\documentclass{amsart}

\usepackage{amsmath,amsfonts,amssymb}

\usepackage{aas_macros} 

\newcommand{\TT}{\mathcal T} 
\newcommand{\FF}{\mathcal F} 

\newcommand{\talk}[1]{{\bf #1}}

\title[GRG18, Session A3]{Report on GRG18, Session A3 \\
Mathematical Studies of the Field Equations}
\author[L. Andersson]{Lars Andersson${}^*$}
\email{laan@aei.mpg.de} 
\thanks{${}^*$ Supported in part by the NSF, under 
contract no. DMS 0407732.}
\address{Albert Einstein Institute, Am M\"uhlenberg 1, D-14476 Potsdam,
  Germany \and
Department of Mathematics, University of Miami, Coral Gables, FL
33124, USA}

\begin{document}

\begin{abstract}
In this report I will give a summary of some of the main topics covered in
Session A3, mathematical studies of the field equations, at GRG18, Sydney.  
Unfortunately, due
to length constraints, some of the topics covered at the session will be very
briefly mentioned or left out altogether. The summary is mainly based on
extended abstracts submitted by the speakers and some of those who presented
posters at the session. I would like to thank all participants for their
contributions and help with this summary.
\end{abstract} 

\maketitle

%\tableofcontents 

%\section{Introduction}

%\mnote{add reference to list of talks and abstract book} 

%The talks in session A3 at GRG18 on mathematical studies of the field
%  equations spanned a wide range of topics and here I will be able to
%  discuss only a few of the talks in any detail. 

%discussed: Andreasson, Uggla, Lim, Nolan, Reula

\section{The Buchdahl inequality} 
%Sharp bounds on $2m/r$ of general spherically symmetric static
%objects}

The Buchdahl inequality, which is included in most textbooks on general
relativity, states that for a static, self-gravitating body,   
\begin{equation} \label{eq:buchdahl} 
2M/R\leq 8/9, 
\end{equation}
where $M$ is the ADM mass and $R$ the area radius of the
boundary of the static body. 
%
%gives a lower bound on the area radius of $R$ of the boundary 
%a static self-gravitating body in terms of the ADM mass. 
%As a consequence,
%inequality also bounds the surface redshift of a static, self-gravitating
%body. 
%
%The Buchdahl inequality has been known to hold only under certain
%conditions, including restrictions on the equation of state. In recent work,
%which he presented at GRG18, H{\aa}kan Andr\'easson has been able to prove
%the Buchdahl inequality with few extra assumptions. 
The proof by Buchdahl, cf. \cite{Buchdahl:1959} of (\ref{eq:buchdahl}) 
assumed that 
%required the assumptions 
%
%In 1959 Buchdahl obtained the bound $2M/R\leq 8/9$ under 
%the assumptions
%that 
the energy density is non-increasing outwards and that the pressure
is isotropic. 
%Here $M$ is the ADM mass and $R$ the area radius of the
%boundary of the static body. 
%The Buchdahl inequality is included in most
%textbooks in general relativity. 
A bound on $2M/R$ has an immediate
observational consequence: if $2M/R<8/9$ then the gravitational red shift
is less than 2 but if $2M/R$ approaches 1 the red shift is unbounded. The
assumptions used to derive the inequality are very restrictive, and as
e.g. pointed out by Guven and \'{O} Murchhada \cite{GM} neither of them
hold in a simple soap bubble and they do not approximate any known
topologically stable field configuration.

In addition to the restrictions implied by the hypotheses made by Buchdahl
there are two other disadvantages associated with this inequality: it
refers to the boundary of the body (i.e. the interior is excluded) and the
solution which gives equality in the inequality is the Schwarzschild
interior solution which has constant energy density and for which the
pressure blows up at the centre. In particular it violates the dominant
energy condition.

\talk{H{\aa}kan Andr\'easson} explained recent work 
\cite{An1}, where 
all four restrictions described above are
eliminated. He considered matter models for which the energy density
$\rho$ and the radial pressure $p$ is non-negative and which satisfies
$p+2p_T\leq\Omega\rho,$ where $\Omega$ is a non-negative constant and
$p_T$ is the tangential pressure, and showed that
\begin{equation}\label{eq:hakan} 
\sup_{r>0}\frac{2m(r)}{r}\leq\frac{(1+2\Omega)^2-1}{(1+2\Omega)^2}.
\end{equation} 
Here $m$ is the quasi-local mass so that $M=m(R).$
The case $\Omega = 1$ gives the bound $2m/r \leq 8/9$. Among the matter
models which satisfy the conditions stated are Vlasov matter and matter
which satisfies the dominant energy condition and has non-negative pressure. 
%
%The conditions on the matter model are very general. Vlasov matter
%satisfies the conditions with $\Omega=1$ and it is sometimes claimed
%\cite{Bo} that a realistic model should satisfy these conditions with
%$\Omega=1.$ Note that this case gives the bound $2m/r\leq 8/9.$ Moreover,
%any matter model with a non-negative pressure which satisfies the DEC
%satisfies the conditions with $\Omega=3.$
%
Inequality (\ref{eq:hakan}) is sharp in the sense of
measures and there are examples
which come arbitrarily close to saturating the inequality. 
Andr\'easson has recently generalized the inequality to the case of 
charged spheres \cite{andreasson:charged}.

\section{Initial-boundary value problems} 
A long standing problem in the analytical and numerical study of the Einstein
equations is the choice of boundary conditions for boundaries near
``infinity''. As is well known, the Einstein equations in harmonic coordinates reduces to a quasilinear 
system of wave equations. It was proved in previous work by Kreiss and Winicour
cf. \cite{kreiss:winicour} 
that the initial-boundary value problem is well posed for 
such systems with the Sommerfeld outgoing boundary condition. The proof
relied on pseudo-differential operator techniques. 

\talk{Oscar Reula} 
reported on a recent proof \cite{kreiss-2007} of this fact which makes use only of partial
integrations. The proof is much more transparent than the previous proof, and
the argument based on partial integration 
is well adapted to the analysis of discretizations of the wave equation 
based on difference operators which 
have good ``partial summation'' properties.

Let $(M,g_{ab})$ be a spacetime with timelike boundary $\TT$. Let $M$ be 
foliated by spacelike surfaces $\Sigma_t$, with timelike unit normal
$n^a$. 
Consider the initial boundary value problem for the wave equation $g^{ab}
\nabla_a \nabla_b \phi = F$ with boundary 
condition of the form $(T^b + a N^b) \nabla_b \phi \big{|}_{\TT} = q$, $a >
0$, where
$T^a$ is a timelike unit vector field which is tangent to $\TT$ 
and $N^a$ is the outward pointing normal to the boundary 
$\partial \Sigma_t = \Sigma_t \cap \TT$ 
in $\Sigma_t$. The Sommerfeld condition is given by the choice $a=1$, which
corresponds to a condition on the derivative of $\phi$ in an outgoing null
direction. 
The proof proceeds by considering the fluxes of energy defined with respect to
a vector field
$u^a= T^a + \delta N^a$, where $\delta > 0$ is a parameter to be chosen.  
Let $\Theta^a{}_b= \nabla^a \phi \nabla_b \phi - \frac{1}{2} \delta^a{}_b \nabla_c
\phi \nabla^c \phi$ be the energy-momentum tensor for the scalar
field. The energy $E$ and outward flux  $\FF$ are defined by integration over
$\Sigma_t$ and $\partial \Sigma_t$, with respect to the
densities $u^b \Theta^a{}_b n_a$ and $u^b \Theta^a{}_b N_a$,
respectively. Using integration by parts one derives by a straightforward
argument an energy estimate bounding $\partial_t E$ in terms of $E$, $F$ and
$\FF$. Expanding out the flux density $u^b \Theta^a{}_b N_a$ and using the
boundary condition to eliminate terms, one finds that for an appropriate 
choice of $\delta = \delta(a) > 0$, 
it contains a negative
multiple of the square gradient of $\phi$ on $\partial \Sigma$ plus terms
which can be estimated in terms of $E, F, \FF$. 
Inserting this in the above mentioned energy estimate yields a form of the
energy estimate which gives the boundary regularity required for
well-posedness. Based on these ideas, the proof 
of well-posedness for the initial boundary can now be completed along
standard lines.

%\section{Moving punctures are not ``Punctures''} 
\section{Moving punctures and black holes}

\newcommand{\beq}{\begin{equation}}
\newcommand{\eeq}{\end{equation}}
\newcommand{\bea}{\begin{eqnarray}}
\newcommand{\eea}{\end{eqnarray}}

\newcommand{\lie}{\mathcal{L}}

%\author{Sascha Husa} \affiliation{Theoretical Physics Institute, University of Jena, 07743 Jena, Germany}
%\author{Mark Hannam} \affiliation{Theoretical Physics Institute, University of Jena, 07743 Jena, Germany}
%\author{Denis Pollney} \affiliation{Max-Planck-Institut f\"ur
%Gravitationsphysik, Albert-Einstein-Institut, Am M\"uhlenberg 1, 14476 Golm, Germany}
%\author{Bernd Br\"ugmann} \affiliation{Theoretical Physics Institute, University of Jena, 07743 Jena, Germany}
%\author{Niall \'O~Murchadha} \affiliation{Physics Department, University
%  College Cork, Ireland}
%\author{Frank Ohme} \affiliation{Theoretical Physics Institute, University of Jena, 07743 Jena, Germany}

%\begin{abstract}
During the time since GRG17, Dublin 2004, there have been tremendous advances 
in numerical relativity, in particular in the binary black hole problem. The
first breakthrough was the work by Frans Pretorius, using a constraint
stabilized harmonic coordinate formulation, with excision. Using this code he was during the
spring of 2005 able to evolve stably several orbits of a binary black hole 
system. This was followed shortly thereafter by announcements from 
several groups using different methods,  that they were able to achieve
similar results. One of the methods which has emerged as a successful
alternative to the harmonic formulation is based on the 
Baumgarte-Shapiro-Shibata-Nakamura (BSSN)
formulation of the Einstein evolution equations, cf. 
\cite{Shibata:1995we,Baumgarte:1998te}, see also \cite{Sarbach:2002bt}. 
%This formulation has been heavily used in BBH work. 
Due
to various issues arising from a combination of several factors (gauge
choice, excision etc.), codes based on BSSN initially suffered from 
instabilities which prevented more than partial orbits to be calculated. 
However when the proper techniques are applied one may 
successfully use the BSSN formulation to 
stably evolve black hole spacetimes, even without excision. The black holes
(BHs) 
are thus included in the domain of computation, and one effectively
treats the BHs as essentially weak singularities in the
Cauchy slices. This approach is now one of the most popular in the numerical
GR community. 
In spite of the success in using this formulation, the reason why it
actually works has remained mysterious. 

\talk{Sascha Husa} 
discussed recent work, cf. \cite{husa:etal,husa:etal:2}, 
which sheds light on this issue. 
%
% Long-term stable black-hole binary simulations have suddenly become
% possible. The most popular method (``moving punctures'') 
%
%
The moving puncture method starts with 
 BHs represented by ``filling'' their interiors with
 wormholes, reduced to singular (``puncture'') points by spatial
 compactification. The moving-puncture method
 \cite{Campanelli:2005dd,Baker:2005vv} is only a minor modification of
 previous approaches, in that no attempt is made to factor out the singular puncture
 geometry during the simulation, but surprisingly in numerical evolutions of a
 Schwarzschild BH using the BAM code \cite{husa:etal:2}, Husa et al. 
found that this leads to a
 drastic change in the geometry 
 of the time slices \cite{husa:etal}. They have  
constructed an analytic solution for the stationary
 state of a nonspinning BH that matches the moving-puncture gauge, and found 
 that the numerical data asymptote to this solution. 

The key result about the geometry of moving punctures is that for a
Schwarzschild black hole the numerical evolutions approach a
stationary slice that neither reaches an internal asymptotically flat
end nor hits the physical singularity, as might be expected for a
stationary slice with non-negative lapse. Rather,
the slice ends at a throat at finite Schwarzschild radius ($R_0 \approx
1.31M$), but infinite 
proper distance from the apparent horizon. This changes the
singularity structure of the ``puncture''. 
It is still a puncture in
that there is a coordinate singularity at a single point in the
numerical coordinates, but it does not correspond to an asymptotically
flat end. In the course of Schwarzschild evolutions Husa et al. 
have found that the
throat does collapse to the origin. Where one would have
expected an inner and an outer horizon, they find only one zero in the
norm of $(\frac{\partial}{\partial t})^a$, corresponding to the outer
horizon. An under-resolved region does develop in the
spacetime (it is the region between the throat and the interior
spacelike infinity), but we are pushed out of causal contact with it. 
The throat itself has receded to infinite proper
distance from the outer horizon. Matter fields or gravitational radiation will be trapped
between the inner horizon and the throat, because unlike the gauge their
propagation is limited by the speed of light; Husa et al. will consider this 
issue in future work.

%Our results suggest several directions for future research directly
%relevant to the black-hole binary problem, such as perturbations of
%the stationary solutions (including constraint violating perturbations
%to check constraint stability of evolution systems), the
%clarification of numerical issues at the discontinuities, 
%and the construction of initial
%data adapted to stationarity, e.g.\ of asymptotically
%cylindrical data.  In the Schwarzschild case local 
%properties of the stationary solution allow one to directly read off spacetime 
%properties from a numerical solution, e.g., the puncture 
%value of $K$ determines the mass; an extension to two moving, spinning black holes
%would be very valuable in numerical evolutions.

%\end{abstract}

%\maketitle

\section{Isolated systems} 
The analysis of the asymptotic structure of asymptotically flat spacetimes is
of central importance in mathematical general relativity. The conformal
compactification setting introduced by Penrose allows one to give a stringent
analysis of the asymptotic properties of spacetimes, including conserved
quantities. Further, the weak cosmic censorship conjecture has a clear
formulation in this setting. The work of Klainerman et
al. \cite{Chr:Kl,Kl:Nic} shows
that asymptotic fall-off conditions on Cauchy data can be introduced so that
the Cauchy development has a conformal compactification with any finite
regularity. It is known that there are large classes of
spacetimes which have a conformal compactification which is regular at
spatial infinity. It is an interesting open question whether the assumption
of smooth null infinity implies some type of rigidity for the spacetime
\cite{corvino:schoen,Chrusciel:2002vb}. The
regular conformal field equations of Friedrich give a clear cut formulation
of this problem, and there is good hope of eventually finding sharp
conditions for regularity. 

\subsection{Stability of Minkowski space} 
The talk of \talk{Lydia Bieri} addressed 
the global, nonlinear stability of solutions of the
Einstein equations in General Relativity. In particular, she discussed
results on the
initial value problem for the Einstein vacuum equations, which generalizes the
results of Christodoulou and Klainerman 
\cite{Chr:Kl}. Every strongly asymptotically flat,
maximal, initial data which is globally close to the trivial data gives rise
to a solution which is a complete spacetime tending to the Minkowski
spacetime at infinity along any geodesic. In \cite{bieri:2}, Bieri has
considered the Cauchy 
problem with more general, asymptotically flat initial data. In particular,
under the assumptions in \cite{bieri:2}, the spacetime curvature fails to be
continuous. In order to show decay
of the spacetime curvature and the corresponding geometrical quantities, the
Einstein equations are decomposed with respect to adequate foliations of the
spacetime. 
%In this work, the main proof is based on a bootstrap argument. 
In the proof of Bieri, one encounters 
borderline estimates for certain quantities with
respect to decay, indicating that the conditions 
%in our main theorem 
concerning 
decay at infinity imposed on the initial data are sharp.

\subsection{Rigidity for asymptotically simple spacetimes} 
\talk{Juan Antonio Valiente Kroon} reported on progress \cite{Val04a} towards
a proof of the following conjecture
concerning asymptotically simple spacetimes: 
{\em 
If an asymptotically
Euclidean, conformally flat, time symmetric initial data set for the
Einstein vacuum equations gives rise to a development admitting a
smooth conformal compactification at null infinity, then the initial
data must be Schwarzschildean in a neighbourhood of infinity. 
}
It should be
noted that the Schwarzschild spacetime is the only stationary spacetime
admitting conformally flat slices. There are further indications of
generalisations of this conjecture for more general classes of data, cf.
 \cite{Val04e,Val05a}. A possible proof of the above
conjecture requires a detailed understanding of the structure of
certain asymptotic expansions of the development of the initial data
near null and spatial infinity. The tools to obtain these expansions
are the ``extended conformal Einstein equations'' and the representation
of spatial infinity known as ``the cylinder at spatial infinity'' which
have been introduced by Friedrich in \cite{Fri98a}. The asymptotic
expansions are obtained by solving, for a given initial data set, 
a hierarchy of interior equations at the cylinder at spatial
infinity. These interior equations allows to transport initial data
from the initial hypersurface up to the ``critical sets'' where null
infinity ``touches'' spatial infinity. The structure of the interior
equations suggests that their solutions should contain very specific
types of logarithmic divergences at the critical sets unless the
initial data is Schwarzschildean. The explicit existence of these
obstructions has been shown in \cite{Val04a}. 

%Valiente Kroon is using the experience
%gained from analysing an analogous problem for the Maxwell field
%propagating on the Schwarzschild spacetime, see \cite{Val07}, to 
%investigate general properties of the solutions to the interior
%equations. 
%
%The preliminary results are encouraging and this work may led to 
%an implementation of a proof of the rigidity
%conjecture by means of an inductive argument. 

\section{Singularities} 

\subsection{Stochastic aspects of generic singularities} 
The proposal of Belinski\v{\i}, Khalatnikov and Lifshitz~\cite{BKL,BKL2,BKL3} (BKL)
on the structure of generic 
cosmological singularities states roughly that the essential properties of the 
asymptotic dynamics of the gravitational field along
typical timelines can be understood by considering certain spatially
homogeneous models. Writing the Einstein equations in terms
of Hubble normalized 
scale invariant frame variables due to Uggla et al. \cite{Ugly,Ugly2,Ugly3} or the
approach of Damour et al. \cite{damour:etal}, based on Iwasawa decomposition, 
each of which relies on a long history
of previous formulations, gives a description of the asymptotics of the
gravitational field in terms of billiard dynamical systems. In the case of
the Hubble normalized formulation, one gets a billiard in the Kasner plane,
cf. \cite{ellis:wainwright}, while for the Iwasawa formulation one gets for the case of
3+1 gravity, a billiard in a domain of hyperbolic space, which is analogous
to the Misner-Chitre billiard.

\talk{Claes Uggla}
presented work
%from a joint paper 
%with Mark Heinzle and Niklas R\"ohr 
\cite{Uggetal}, building on
\cite{Ugly,Ugly2,Ugly3}, 
which generalizes and makes rigorous 
some aspects of the previous work of BKL and others
%concerning generic spacelike singularities in general
%relativity starting from the Iwasawa based Hamiltonian formulation of Nicolai
%et al. The work of Uggla et al. focuses 
on the stochastic aspects of the
system which arise due to the chaotic nature of the asymptotic billiard
systems. Using a combination of dynamical and statistical analyses and in
part heuristic arguments, this work describes the generic properties of a
``billiard attractor''.
% and showed that a number of terms which could in
%general be significant for the dynamics are suppressed due to statistical
%effects. 
%
It turns out that several degrees of freedom, which
a priori could have been of relevance for the asymptotic behavior are,
generically, 
statistically suppressed. The dynamical and statistical
arguments of Uggla et al.
may be contrasted with results concerning asymptotic behavior in Bianchi type
IX obtained using purely dynamical arguments \cite{Ring}.
The results presented by Uggla suggest
that the role of Bianchi type IX as a model for the 
asymptotic dynamics of generic
singularities should be re-evaluated, in view of the fact that 
generic singularities exhibit features that are not shared by Bianchi type
IX. 
%moreover, there are no proofs about some aspects of the asymptotic dynamics
%of Bianchi type IX that presumably are of generic importance.

\subsection{Kinematic and Weyl singularities}
%Speaker: Woei Chet Lim (wlim@princeton.edu)%
%
%Department of Physics, Princeton University, Princeton, NJ 08544, USA.
%%
%
%Ref: \cite{lim:06,lim:07}
For expanding Bianchi cosmologies with a tilted perfect 
fluid with linear equation of state $p=(\gamma-1)\rho$, all timelike 
and null geodesics are complete into the future. However, the fluid congruence 
may be incomplete into the future (called a congruence singularity), 
accompanied by the blow-up of kinematic 
quantities associated with the fluid congruence.
\talk{Woei-Chet Lim} discussed recent work \cite{lim:06,lim:07} on 
the nature of such singularities. 
Much emphasis has been placed on the blow-up of the components of the Weyl 
tensor associated with the fluid congruence. However, as shown by examples
due to Lim et al.,
this phenomenon is independent of the incompleteness of the fluid 
congruence. 
Hence it is necessary to differentiate congruence 
singularities into kinematic
singularities 
and 
Weyl singularities.
In particular, the fluid may become
extremely titled or the kinematic or Weyl components may blow up, depending
on whether $\gamma$ exceeds certain critical values. 
 
%Through examples, we show that generally the perfect fluid becomes 
%extremely tilted into the future if the parameter $\gamma$ exceeds some 
%threshold $\gamma_{\rm tilt}$.
%Similarly, the kinematic and Weyl components blow up if $\gamma$ exceeds 
%some thresholds $\gamma_{\rm sing}$ and $\gamma_{\rm Weyl}$ respectively.
%We also present a neat relation between 
%the limit of the ratio $\max|C_{abcd}|/H^2$
%and whether the threshold $\gamma_{\rm sing}$ is larger than the threshold 
%$\gamma_{\rm Weyl}$.
%This work indicates that it is important to 
%We hope the audience will pay more attention to the blow-up of 
%the kinematic quantities.

\subsection{Perturbations of naked singularity spacetimes} 
%Bounds for scalar waves and linear perturbations in self-similar
%naked singularity space-times.
%
%\vskip12pt Brien C. Nolan
%
%\vskip12pt
The scalar field may be considered as a toy model for perturbations of a
background spacetime. 
\talk{Brien Nolan} 
discussed some rigorous mathematical results that probe the linear
stability of naked singularities in self-similar collapse. He showed
that the multipoles of a minimally coupled massless scalar field
propagating on a spherically symmetric self-similar background
space-time admitting a naked singularity maintain finite $H^{1,2}$
norm as they impinge on the Cauchy horizon. Further, 
each multipole obeys a point-wise bound at the horizon, as does its
locally observed energy density \cite{bcn:06}. The null energy condition is
assumed to hold in the background spacetime.
%%
%
%We do not specify the
%matter content of the space-time, but assume that Einstein's
%equation and the null energy condition hold. 
%The scalar field may be
%considered a toy model for a perturbation of the background space-time. 
In order to view the scalar field as a toy model for perturbations of a
background spacetime, the matter model must be specified.  To study such
perturbations, the matter content must be specified.  Nolan considers  what
in this context is the simplest case: that of null dust - i.e. the stability
of the Cauchy horizon in self-similar Vaidya space-time is studied.  The
results for the scalar field carry over to odd-parity linear perturbations at
the level of both the metric and the curvature \cite{bcn:07a}. For even
parity perturbations, the linearised Einstein equations can be reduced to a
5-dimensional first order symmetric hyperbolic system. The components of the
state vector are gauge invariant metric and matter perturbation
quantities. Nolan shows that the $L^2$ norm of the state vector blows up at
the Cauchy horizon and that solutions for which the $L^\infty$ norm of the
state vector does not blow up at the Cauchy horizon are unstable in the space
of test function initial data. This provides strong evidence that the Cauchy
horizon of the self-similar Vaidya space-time is unstable \cite{bcn:07b}.

\subsection{Spacetime boundaries and properties of singularities} 
In his talk, 
\talk{Benjamin Whale} reported on joint work with Sue Scott, where the
a-boundary and the a-boundary singularity theorem are applied to analyze the 
physical properties of singularities. 

%The singularity theorems of Penrose and Hawking (among others) demonstrate
%that singularities are a general feature of GR, and provide strong evidence
%that, if not singularities, then some extreme behaviour will exist in any
%quantum theory of gravity.  Thus it is of interest to discuss the physical
%properties of these singularities. Until recently, however, we lacked the
%mathematical framework to adequately tackle this question. With the advent of
%the a -boundary (Scott and Szekeres, 1994) and the a - boundary singularity
%theorem (Ashley and Scott, 2003) it is now possible to make clear in-roads
%into this difficult area. In this talk we wi ll outline the a-boundary, the a
%- boundary singularity theorem and describe progress in the quest to reveal
%the physical properties of these problematic peculiarities.

\section{Quasi-local aspects} 

\noindent
\subsection{Towards the quasi-localization of canonical GR}
% (L\'aszl\'o B. Szabados)} 
Perhaps the most natural way of introducing the conserved quantities 
in asymptotically flat spacetimes is the canonical/Hamiltonian one. 
A key observation in the Hamiltonian formulation of GR, due to 
Arnowitt, Deser and Misner \cite{ADM}, is that the evolution parts of 
Einstein's equations can be recovered formally as canonical equations 
of motion in the phase space, in which the constraint function (i.e. 
whose vanishing is just the constraint parts of Einstein's equations) 
play the role of the Hamiltonian. 
However, as Regge and Teitelboim stressed, if we want to recover the 
correct evolution parts of Einstein's equations as partial differential 
equations for {\em smooth fields} rather than some distributional 
generalization of them, then the Hamiltonian in the canonical 
equations of motion must be {\em functionally differentiable} with 
respect to the canonical variables \cite{RT}. Regge and Teitelboim 
showed that such a Hamiltonian can be obtained by adding an 
appropriate 2-surface integral to the constraint functions. The 
remarkable property of this Hamiltonian is that the ADM energy-momentum 
and angular momentum can be recovered as the value of the Hamiltonian 
on the constraint surface with appropriately chosen lapse and shift. 
On the other hand, the investigations of Beig and \'O Murchadha 
showed that for given boundary conditions on the canonical variables 
the asymptotic form of the lapse and the shift is already determined 
if we require that the evolution equations preserve the boundary 
conditions \cite{BM}. Moreover, they also showed that both the 
constraint functions and the Hamiltonians close to Poisson algebras, 
the former being an ideal in the latter, and their quotient, the 
so-called algebra of observables, is isomorphic to the Poincare algebra. 
The ADM conserved quantities then can be considered as appropriate 
coordinates on this algebra of observables. 

\talk{L\'aszl\'o Szabados} considered in his talk  the boundary conditions
for canonical vacuum GR at the quasi-local level, i.e. when the spacelike
hypersurface on which the canonical variables are defined is compact  with
smooth 2-boundary ${\mathcal S}$ \cite{Sz1} . The key ideas used by Szabados
are found by analogy with the Hamiltonian analysis in the asymptotically
flat case  above, namely   the requirement of the functional
differentiability of all the  functions whose Poisson bracket must be
calculated, and  the requirement of the compatibility of the evolution
equations and  the boundary conditions both on the canonical variables and
the lapse  and the shift.  It has been  shown by Szabados  \cite{Sz2} that
fixing the  area element on the 2-surface  ${\mathcal S}$ (rather than the
whole induced 2-metric) is enough to have  a well defined constraint algebra,
a well defined Poisson algebra of  basic Hamiltonians parameterized by lapses
that are vanishing on ${\mathcal  S}$ and shifts that are tangent to   and
divergence free on ${\mathcal  S}$. Their quotient is a Lie algebra of a
class of 2-surface observables.

The evolution equations preserve these boundary conditions, and the
observables (realized as the value of the basic Hamiltonians on the
constraint surface) are 2+2--covariant, gauge-invariant and provide a
representation of the Lie algebra of the divergence-free vector fields  on
${\mathcal S}$. Szabados pointed out, that in special,  standard situations
the observables appear to  behave as angular momentum.

%\bibliography{A3_szabados}

\subsection{The Einstein scalar field at finite infinity}
%William J CLAVERING
%Astronomy Unit, Queen Mary University of London, London, UK
The `finite infinity' paradigm of Ellis was  proposed to study isolated
gravitational systems, in our universe, where we experience the presence of
other local matter fields and a cosmological background \cite{Ellis:2002hn}.
In this context it is not possible to make the assumption of asymptotic
flatness at infinity, under which boundary conditions on matter have been
studied in detail.  Instead, a smooth timelike surface $\mathcal F$, is
introduced. The timelike surface $\mathcal F$ should be located at a
large finite distance from the centre of the local system of interest,
with the aim to study the fields generated from this on $\mathcal F$.

\talk{William Clavering} in his talk examined  the behaviour of the
spherically symmetric Einstein scalar field at $\mathcal F$.  He considered
the evolution of a scalar field over a domain bounded by an initial
hypersurface, characteristic curves, and $\mathcal F$.  Using  Hawking's mass
formula \cite{Hawking:1968qt},  he has studied the consequences of imposing
conditions on the mass-energy flux at $\mathcal F$.  Two examples were
discussed; each a perturbation of a static exact solution.  For a
perturbation of Schwarzschild there is no mass-energy flux of the scalar
field. Clavering conjectures  that this is the case in all vacuum spacetimes.
The analysis has been repeated for the pure scalar field case of the Wyman
spacetime \cite{Wyman:1981bd}.  In this case, non-trivial expressions for the
mass-energy flux in terms of the scalar field at $\mathcal F$ were obtained.
%This work is currently in preparation.

\subsection{Quasi-local energy inside the event horizon}
% (A. Lundgren)}
%
%\title{Quasilocal Energy Inside the Event Horizon}
%
%\author{Andrew P. Lundgren}
%\affiliation{Cornell University}
%
%\author{Bjoern S. Schmekel}
%\affiliation{University of California, Berkeley}
%
%\author{James W. York, Jr.}
%\affiliation{{}Cornell University}
%
%\maketitle
%
Pointlike objects cause many of the divergences that afflict physical
theories.  For instance, the gravitational binding energy of a point particle
in Newtonian mechanics is infinite.  In general relativity, the analog of a
point particle is a black hole and the notion of binding energy must be
replaced by quasilocal energy.  The quasilocal energy (QLE) derived by York,
and elaborated by Brown and York \cite{1993PhRvD..47.1407B}, is finite
outside the horizon but it was not considered how to evaluate it inside the
horizon.  

\talk{Andrew Lundgren} presented 
a prescription for finding the QLE inside a horizon, and
showed that it is finite at the singularity for a variety of types of black
hole. It turns out that the energy is typically concentrated just inside the horizon, not at
the central singularity.
It is impossible to localize gravitational energy due to the equivalence
principle, so it is meaningless to define gravitational energy at a single
point.  This problem is avoided by considering the gravitational energy in a
region of space.  The boundary of the region is a two-dimensional surface,
and in fact the quasilocal energy is defined only in terms of quantities
defined on the enclosing surface (specifically the induced metric and
extrinsic curvature).  The QLE is defined only in terms of the gravitational
degrees of freedom and makes no mention of any other fields that are present.
However, since gravity couples to everything, the QLE counts all energy that
is present in the region.

Lundgren et al. \cite{lundgren:084026} have considered 
only spherically-symmetric static
black holes.  The definition of the QLE in
\cite{1993PhRvD..47.1407B} was extended 
to be valid inside the event horizon, which is
only a coordinate singularity.  In the Schwarzschild case, the QLE at the
central singularity is zero.  The analogous quantity for a classical field of
a point particle diverges at the center.  The gravitational field itself
carries energy which gravitates, causing the nonlinearity of general
relativity. Lundgren remarked that the 
nonlinearity somehow provides a mechanism that cures
the divergence expected of a point particle field. The definition of the QLE
presented by Lundgren can be applied also to the deSitter case and the case
with non-vanishing charge.

\section{Black holes} 

\subsection{Black hole rigidity and spacetimes with compact Cauchy horizon}
\talk{Jim Isenberg} reported on recent work with Vince Moncrief
which studies analytic solutions of Einstein's equations
containing nondegenerate compact Cauchy horizons with non closed
generators. If certain hypotheses are added, 
the spacetimes necessarily admit an isometry which
acts along the generators of the horizon. 
Isenberg showed how to use these results to prove
that stationary (non static) analytic black holes in arbitrary dimensions 
necessarily admit
symmetries which are independent of the presumed stationarity symmetry. This
work is closely related to work of Hollands et al. \cite{Hollands:2006rj}.

\subsection{Electromagnetic field on the background of high dimensional black
  holes} 
%
%\title{Electromagnetic field on the background of high dimensional black holes\\[1ex]
%{\large \it and two no-go theorems for generalizations to higher-dimensional\\[-0.5ex] Pleba\'nski--Demia\'nski metric}}
%
%
%\author{Pavel Krtou\v{s}\\
%\small Institute of Theoretical Physics,\\ 
%\small Faculty of Mathematics and Physics, Charles University,\\ 
%\small V Hole\v{s}ovi\v{c}k\'ach 2, Prague, Czech Republic}
%
%\date{GRG-18, Sydney, July 12, 2007}
%\maketitle
%
In the last years various generalizations of black hole solutions 
to the high dimensional gravity have been found. One of them describes a
generally rotating black hole with NUT charges. This solution possesses 
several nice properties as, e.g., the existence of the Killing-Yano tensor,
the complete integrability of the geodesic motion or the separability of 
the Hamilton-Jacobi and Klein-Gordon equations. 
Its generalization including the 
acceleration of the black holes or the electromagnetic field is not, 
however, straightforward. 

In four dimensions it is possible to generalize the black hole solution to
describe the accelerated black holes (the Pleba\'nski-Demi\'nski solution).
This transition is achieved by the conformal rescaling of the metric
accompanied by a modification of some metric functions. 
\talk{Pavel Krtous} showed in his talk that such a procedure 
is not sufficiently general in higher dimensions---only 
the maximally symmetric spacetimes in `accelerated' 
coordinates are obtained.
Further, he presented general algebraically special solutions of 
the Maxwell equations on the background of the mentioned high dimensional 
generally rotating black hole. They are adjusted to the algebraic structure 
of the background which is reflected by the existence of the principal 
Killing-Yano tensor. Such electromagnetic fields generalize the field of 
charged black holes in four dimensions. 
However, one may show that in higher dimensions such fields cannot be 
easily modified in such a way that they would satisfy full Maxwell-Einstein equations.

%################# Equation commands #########################

\newcommand{\eqn}[2]{ \begin{equation*} #2  \end{equation*} }
\newcommand{\gath}[2]{ \begin{gather*} #2 \end{gather*} }
\newcommand{\lalign}[2]{ \begin{align} #2 \end{align} }
\newcommand{\alin}[2]{ \begin{align*} #2 \end{align*} }
\newcommand{\leqn}[2]{\begin{equation} #2 \label{#1} \end{equation}}
\newcommand{\leqnarr}[2]{ \begin{eqnarray} #2 \end{eqnarray} }
\newcommand{\lgath}[2]{ \begin{gather} #2 \end{gather} }
\newcommand{\ltag}[1]{ \label{#1} }
\newcommand{\ntag}[1]{ \tag{{\scriptsize#1}} \label{#1} }
\newcommand{\nnotag}{ \notag }
\newcommand{\sect}[2]{\section{#2}\label{#1}}
\newcommand{\subsect}[2]{\subsection{#2}\label{#1}}

\newcommand{\ep}{\epsilon} 
\newcommand{\del}{\partial} 

\section{Post-Newtonian expansions} 
Post-Newtonian expansions (PNEs) are expansions of the Einstein equations in the
parameter $1/c$, around $c=\infty$, where $c$ is lightspeed. The limit
$c=\infty$ is the Newtonian limit of general relativity. 
%Post-Newtonian expansions have 
%This problem has
Such expansions have 
been studied since the discovery of general relativity and there is a large
literature available. 
% by many
%people and there is a large number of results available in the
%literature. 
%PNEs 
%Post-Newtonian
%expansions 
%are of fundamental importance in 
%many applications of 
%areas of
%experimental gravity . 
%PNEs are of fundamental importance in In particular, such 
%PNEs have been used eg.  
%in the analysis of 
%to test
%alternatives to Einstein gravity, for example using the 
%observations of the
%binary pulsar NNN. 
Recently, post-Newtonian expansions have been used to
calculate wave forms for the gravitational wave emissions from binary black
hole inspirals. These have been compared with numerically calculated wave
forms and turn out to be highly accurate until the last stage of the inspiral. 
%It is an interesting problem to prove rigorously the correctness of 
%post-Newtonian expansions. 

The majority of results on PNEs are based on formal expansions
which are used to calculate 
approximate values of physical quantities, see eg. \cite{Ehl86,BFN05} and
references therein. 
However, this formal approach does not deal with the question of convergence.
%
%The main difficulty with the
%formal expansions is that they leave completely unanswered the
%question of convergence. 
In the absence of a precise notion of
convergence, it becomes unclear to what extent the formal
expansions actually approximate relativistic solutions.
In view of the importance of the applications of PNEs, it is interesting to
give a rigorous foundation to the procedure. 
% and also 
%to investigate
%the behavior of the gravitational and matter fields for $c$ close to
%$\infty$. 
%$\ep$ in
%a neighborhood of zero. 
%For some classic and
%recent
%results of this type see 
%\cite{BD86,BFN05,Dau64,DBKM,Chand65,EIH38,Ehl86,Kun72,Kun76,KD,PW00,Will05} and
%\cite{NNN} and 
%references cited therein.
%

\talk{Todd Oliynyk} discussed his recent work on post-Newtonian expansions
for the Einstein-Euler equations.
%, i.e. the system
%The Einstein-Euler equations are given by the following system of equations
%\lgath{EEeqn}{
%G^{ij} = \frac{8\pi G}{c^4} T^{ij} \label{EEeqn.1} \\
%\nabla_{i} T^{ij} = 0 \label{EEeqn.2} } where the stress-energy
%tensor for the fluid is given by \lgath{EEdefs}{ T^{ij} = (\rho +
%c^{-2} p)v^i v^j + p g^{ij} \,  } with $\rho$ the fluid density,
%$p$ the fluid pressure, and $v$ the fluid four-velocity normalized
%by $v^iv_i = -c^2$, $c$ the speed of light, and $G$ the
%Newtonian gravitational constant. 
Oliynyk considers expansions of solutions to
these equations in the parameter 
$\ep = v_T/c$ about 
$\ep = 0$, where
$v_T$ is a characteristic velocity scale associated with the
fluid matter. 
%, are known as \emph{post-Newtonian expansions}. 
By rescaling the gravitational and matter variables, the
Einstein-Euler equations can be written as
\leqn{EEhateqns.intro}{ G^{ij} = 2\kappa \ep^4 T^{ij} \quad
\text{and} \quad \nabla_{i} T^{ij} = 0 } where
$\kappa$  $=$ $4\pi G\rho_T /v_T^2$, $v_i v^i$ $=$ $-\ep^{-2}$,
$\rho_T$  is
a characteristic value for the fluid density, and
$t$ $=$ $x^4/v_T$ is a ``Newtonian'' time coordinate.
%
%NNNN%
%
%One expects that there exists a class of solutions to the 
%Einstein-Euler equations \eqref{EEhateqns.intro} that admit expansions in $\ep$, at least to low orders,
%where the $0^\text{th}$ order expansion satisfies the 
%Poisson-Euler equations
%\lalign{newtB.intro}{
%\del_t \rho + \del_I(\rho w^I) & = 0 \, , && (I,J=1,2,3)\label{newtB.1.intro}\\
%\rho(\del_t w^J + w^I\del_I w^J) & =
%-(\rho\del^J\Phi + \del^J p) \, , && (\del^I = \delta^{IJ}\del_J )\label{newtB.2.intro} \\
%\Delta \Phi &=   \rho \, , && (\Delta = \del_I \del^I) \label{newtB.3.intro} \,
% }
%of Newtonian gravity. 
%As above, $\rho$ and $p$ are the fluid density and pressure,
%respectively, while $w^I$ is the fluid (three) velocity.
%%
%%
%
%NNNN
%
In his talk, Oliynyk presented results on PNEs for a class of polytropes, 
which go beyond formal
considerations \cite{Oli05e,Oli05d}.
By introducing suitable renormalized
variables for which the limit $\ep \to 0$ makes sense, and introducing a
suitable gauge, Oliynyk derives a family of symmetric hyperbolic systems,
depending on the parameter $\ep$. This system is studied in a class of
weighted Sobolev spaces $H^k_{\delta,\ep}$, which interpolate between the
weighted spaces $H^k_\delta$ and the standard spaces $H^k$. Using these
spaces, it is possible to prove $\ep$-dependent energy estimates for
solutions to the Einstein-Euler equations. These estimates are then used to
prove the existence of convergent expansions in $\ep$ for suitably chosen initial data. 
Oliynyk discussed the relation between his convergent expansions and the
formal PNEs. In order to recover the PNEs to a certain order requires the
initial data to be chosen correctly. The method used to choose initial data
is based on ideas in
\cite{BK}. 
Oliynyk discussed how to recoved the post-Newtonian expansion to 
$2^{\text{nd}}$ order 
in his framework. 

\section{Miscellaneous} 

\subsection*{Wave tails in curved spacetimes}  
\talk{Risto Tammelo} reported on joint work with Laas and Mankinen  dealing
with the wave tails of scalar and electromagnetic fields.  Their work shows
that the intensity of
the tail of the electromagnetic wave pulse emitted by a wave source within a
compact binary in the vicinity of the geometric shadow of the source of
gravitation can be of the same magnitude as the intensity of the direct
pulse. The energy carried away by the tail can amount to approximately 10\% of
the energy of the low-frequency modes of the direct pulse.  
As an example of
an exactly solvable model system, a scalar wave field on a particular
Robertson-Walker spacetime has been considered. 
In the case of minimal coupling, if
the metric describes the Friedman dust-dominated universe, the higher-order
multipole solutions do not contain a wave tail term. A massless
nonconformally-coupled scalar field is also considered in a class of
Robertson-Walker spacetimes. 
%The retarded and advanced Green's functions are
%calculated analytically and applied to find the Hadamard fundamental solution
%in terms of a hypergeometric function in some special cases. By applying a
%particular coordinate transformation and allowing unrestricted growth of the
%coupling constant of the scalar and gravitational fields, 
Tammelo et al. show 
that an initially massless scalar field in the Robertson-Walker universe can
obtain a mass in the corresponding asymptotic Minkowski space region.

\subsection*{Ricci flow and the Einstein equations} 
\talk{Eric Woolgar} described work 
on the Ricci flow
of asymptotically flat manifolds in the rotationally symmetric case
\cite{OW}. 
This paper shows that if no 
minimal hypersphere is present initially, then one never
forms.  
The flow then exists for all future 
time and converges to flat spacetime in the limit of infinite time, the 
limit being taken in the Cheeger-Gromov sense. The mass does not change 
during the flow, but the quasilocal mass within any fixed hypersphere 
about the origin (defined either by fixing the proper radius, the surface 
area, or the enclosed volume) evaporates away smoothly.
\talk{Mohammad Akbar} discussed relations between Ricci solitons and
solutions to the Einstein-scalar field equations. 

\subsection*{Posters}
Several interesting abstracts had to be relegated to the poster
session. Among these were abstracts by \talk{H{\aa}kan
  Andr\'easson} 
%based on joint work with Kunze and Rein 
on the formation of black holes in spherically symmetric
gravitational collapse
\cite{AnRe}, \talk{Jim Isenberg} on the AVTD behavior of $T^2$ symmetric
solutions of the Einstein vacuum equations, 
\talk{Mak\-o\-to Narita} on cylindrically symmetric expanding
spacetimes,  
\talk{Oscar Reula}, 
%based on joint work with Kozameh and Kreiss 
\cite{kozameh-2007}, 
showing that
the Robinson Trautman Maxwell equations do not constitute a well posed initial
value problem, and 
\talk{Juan Antonio Valiente
  Kroon} 
%based on joint work with Garcia-Parrado Gomez-Lobo 
on a characterization of Schwarzschild initial data sets \cite{VK:schw}.

%\talk{Kentaro Yasuno} considered the

%\bibliographystyle{amsplain}
%\bibliography{a3sum}

\providecommand{\bysame}{\leavevmode\hbox to3em{\hrulefill}\thinspace}
\providecommand{\MR}{\relax\ifhmode\unskip\space\fi MR }
\providecommand{\MRhref}[2]{%
  \href{http://www.ams.org/mathscinet-getitem?mr=#1}{#2}
}
\providecommand{\href}[2]{#2}

\end{document}